\documentclass[usenatbib]{mn2e}
\usepackage{natbib,epsfig}

\catcode`\@=11
\def\gta{\ifmmode{\,\mathrel{\mathpalette\@versim>\,}}
    \else{$\,\mathrel{\mathpalette\@versim>}\,$}\fi}
\def\lta{\ifmmode{\,\mathrel{\mathpalette\@versim<\,}}
    \else{$\,\mathrel{\mathpalette\@versim<}\,$}\fi}
\def\@versim#1#2{\lower 2.9truept \vbox{\baselineskip 0pt \lineskip
    0.5truept \ialign{$\m@th#1\hfil##\hfil$\crcr#2\crcr\sim\crcr}}}
\catcode`\@=12  

\usepackage{graphics} \usepackage{amssymb}
\usepackage{setspace} 

\newcommand{\beq}{\begin{equation}}
\newcommand{\eeq}{\end{equation}}

\newcommand{\figref}[1]{Fig.~\ref{#1}}
\newcommand{\fig}[1]{Figure \ref{#1}}

\newcommand {\hi} {{\rm H}\,{\small\rm I}}

\newcommand {\ovi} {{\rm O}\,{\small\rm VI}\,}
\newcommand {\feh} {\hbox{[Fe/H]}}
\newcommand {\acrit} {\alpha_{\rm crit}}
\newcommand {\gsim}{\,\lower.7ex\hbox{$\;\stackrel{\textstyle>}{\sim}\;$}}
\newcommand {\lsim}{\,\lower.7ex\hbox{$\;\stackrel{\textstyle<}{\sim}\;$}}

\newcommand {\rev}[1] {#1}

\def\yr{\,{\rm yr}}\def\Myr{\,{\rm Myr}}\def\Gyr{\,{\rm Gyr}}
\def\msun{\,M_\odot}
\def\msunyr{\,M_\odot \yr^{-1}}
\def\K{\,{\rm K}}

\def\c{{\rm c}}\def\w{{\rm w}}\def\h{{\rm h}}\def\m{{\rm m}}
\def\e{{\rm e}}
\def\kpc{\,{\rm kpc}}\def\pc{\,{\rm pc}}
\def\kms{\,{\rm km\,s}^{-1}}
\def\cm{\,{\rm cm}}

\title[Gas accretion onto star-forming galaxies]
{The mode of gas accretion onto star-forming galaxies}

\author[F. Marinacci et al.]
{Federico Marinacci$^{1,2}$, James Binney$^{2}$,  Filippo
Fraternali$^1$,\newauthor Carlo Nipoti$^1$, Luca Ciotti$^1$, and Pasquale Londrillo$^3$\\
$^{1}$Department of Astronomy, University of Bologna, via Ranzani 1, 40127,
Bologna, Italy\\ 
$^{2}$Rudolf Peierls Centre for Theoretical Physics, Oxford University, Keble Road, Oxford OX1
3NP, UK\\
$^{3}$INAF-Osservatorio Astronomico di Bologna, via Ranzani 1, 40127,
Bologna, Italy}

\begin{document}

\date{Accepted 2010 January 14.  Received 2009 December 21; in original form 2009 October 27}

\pagerange{\pageref{firstpage}--\pageref{lastpage}} \pubyear{2009}

\maketitle

\label{firstpage}

\begin{abstract}
It is argued that galaxies like ours sustain their star formation by
transferring gas from an extensive corona to the star-forming disc. The
transfer is effected by the galactic fountain --  cool clouds that
are shot up from the plane to kiloparsec heights above the plane. The
Kelvin-Helmholtz instability strips gas from these clouds. If the pressure
and the the metallicity of the corona are high enough, the stripped gas causes a
similar mass of coronal gas to condense in the cloud's wake. Hydrodynamical
simulations of cloud-corona interaction are presented. These confirm the
existence of a critical ablation rate above which the corona is condensed,
and imply that for the likely parameters of the Galactic corona this rate
lies near the actual ablation rate of clouds. In external galaxies trails of
\hi\ behind individual clouds will not be detectable, although the integrated
emission from all such trails should be significant. Parts of the trails of
the clouds that make up the Galaxy's fountain should be observable and may
account for features in targeted 21-cm observations of individual
high-velocity clouds and  surveys of Galactic \hi\ emission. Taken in
conjunction with the known decline in the availability of cold infall with
increasing cosmic time and halo mass, the proposed mechanism offers a
promising explanation of the division of galaxies between  the blue
cloud to the red sequence in the colour-luminosity plane.

\end{abstract}


\section{Introduction}

A considerable body of evidence from diverse sources leads to the conclusion
that star-forming disc galaxies such as the Milky Way accrete $\gta1\msun$ of
gas each year \citep[e.g.][and references therein]{Pagel97,Chiappini01,Sancisi08}, and have
built up their observed discs gradually over the last $10\Gyr$
\citep[e.g.][]{Twarog80,Chiappini01,AumerB}. What remains unclear is from what reservoir this gas is
drawn, and how it enters the thin gas disc within which stars are formed. 

A central question is the temperature of the reservoir: is this low enough
($T\lta10^4\K$) for the reservoir to contain largely neutral gas, or
comparable to the virial temperature, $T\gta10^6\K$, of the gravitationally
bound groups within which Milky-Way type galaxies currently reside?

In recent years there has been much enthusiasm for so-called cold-mode
accretion of gas that has failed to be shock heated to the virial temperature
\citep{birnb03,Binney04,Keres05,Cattaneo06,Keres09,Hopkins09,Law09}.  Cosmological
simulations suggest that cold-mode accretion is the dominant process at
redshifts $z\gta2$, but gradually becomes less important. A powerful argument
against its currently being the dominant process is the persistent failure of
21-cm surveys to identify significant bodies of intergalactic \hi\ in the
nearby Universe \citep{SargentLo,Pisano04,Kovac09}. In particular the
Galaxy's ``high-velocity'' \hi\ clouds, which as late as 1999 were argued to
be distant and massive \citep{BlitzS}, are now known to be at distances
$\sim10\kpc$ and have masses $\lta10^5\msun$ too small for these clouds to be
a cosmologically significant reservoir of gas \citep{Pisano04,Wakker08,Sancisi08}.

\cite{Spitzer56} already inferred from the presence of interstellar
absorption in the spectra of high-latitude stars that the Galactic disc must
be embedded in pervasive medium of temperature $\sim10^6\K$. The empirical
case for such ``coronal gas'' was greatly strengthened by the Copernicus and
FUSE missions, which detected ions such as \ovi on high-latitude sight lines
to distant UV sources, in particular sight lines that pass close to
high-latitude \hi\ clouds \citep{SpitzerJenkins,Sembach03}. The highly
ionised gas detected must be collisionally ionised and is most readily
interpreted as material at the interface between coronal gas with $T>10^6\K$
and clouds of cooler, partly neutral \hi.

Cosmology strongly suggests that galaxies should be embedded in
virial-temperature coronae. First, standard cosmology predicts that only a
minority of baryons are contained in stars and cool interstellar gas
\citep[e.g.][]{FukugitaP04,Komatsu09}. In rich
clusters of galaxies the ``missing'' baryons are directly detected through
their X-ray emission \citep{Sarazin}.  In lower-density environments, such as the Local
Group, it is thought that the surface-brightness of X-ray emission is too low
to be detected by current instrumentation \citep{Rasmussen09}. Second, three
lines of argument indicate that star-formation is an inefficient process in
which as much gas is ejected from a star-bursting system as is converted into
stars: (i) we actually see winds blowing off star-forming discs
\citep{CohenBlandH,M82}; (ii) the spectra of several quasars show
blue-shifted absorption-line systems indicative of massive winds flowing away
from the star-bursting host galaxy \citep{Pettini,Adelberger03}; (iii) in clusters of
galaxies of order half the metals synthesised by stars are in the
intergalactic medium \citep{Sarazin}.

Our premise in this paper is that most of the baryons originally associated
with the Milky Way's dark matter comprise a corona of gas at the virial
temperature, and that the gas that sustains star formation in the disc is
drawn from this  corona. The question is how gas makes the transition from a
pressure supported corona to the centrifugally supported disc.

For more than thirty years X-ray astronomers have studied the
virial-temperature coronae of rich groups and clusters of galaxies. Most of
these coronae have central cooling times that are significantly shorter than
their ages, and it is natural to ask whether we can understand the
Milky-Way's corona by extrapolating results for cluster ``cooling flows'' to
lower masses. It seems, however that the dynamics of these systems is
qualitatively different from the dynamics of the Milky Way's corona because
the central galaxies of rich clusters do not have massive stellar discs.  The
extent to which gas in rich clusters is cooling (as opposed to radiating) is
controversial, but it is now widely accepted that radiative losses by the
inner corona are largely offset by mechanical feedback from the central black
hole \citep{BinneyT,OmmaB,NipotiB05,Best07}. It is important to understand the
origin of this qualitative difference in the dynamics of the coronae of
star-forming galaxies of the ``blue cloud'' and ``green valley'' in
colour-luminosity space \citep{Blanton03} and that of the coronae
of massive galaxies within the ``red sequence''.

Sensitive \hi\ observations of star-forming disc galaxies reveal that these
galaxies keep  $10$ to $25$ per cent of their \hi\ a kiloparsec or
more above or below their disc plane \citep[][and references therein]{Boomsma08,Fraternali09}. Most of
this gas is thought to have been driven out of the disc by supernova-powered
bubbles \citep{ShapiroF76,HouckB90} and must consist of clouds that are
moving on essentially ballistic trajectories because if it were in
hydrostatic equilibrium, its vertical density profile would be very much
steeper than that observed by virtue of its low temperature
\citep{Collins02}. The gas is expected to return to the disc within
$\sim100\Myr$, so the phenomenon of extraplanar \hi\ is evidence for galactic
fountains \citep{Bregman97}. \hi\ observations of extraplanar gas in the
Milky Way are hard to interpret on account of our location within the plane,
but the Leiden--Argentina--Bonn (LAB) survey \citep{LAB}, which mapped local
\hi\ emission with high sensitivity, is consistent with the Milky Way having
a distribution of extraplanar \hi\ that is similar to the distributions of
\hi\ studied around external nearby galaxies.

The extraplanar gas of nearby galaxies has two key properties: (i) the mean
rotation speed of the gas declines quite rapidly with distance from the
plane; (ii) the gas shows net motion towards the galaxy's symmetry axis.
Fraternali \& Binney (2006; hereafter FB06) and Fraternali \& Binney (2008;
hereafter FB08) fitted models of a galactic
fountain to observations of NGC\,891 and NGC\,2403 and showed that these
models can only account for the observed rotation rates and inward motion of
the extraplanar gas if the observed \hi\ clouds accrete gas such that the
mass of a cloud exponentiates on a timescale $\sim650\Myr$. Such an accretion
rate simultaneously accounts for data from the two rather different
galaxies modelled and yields rates of accretion onto the discs,
$\sim3.4\msun\yr^{-1}$ and $0.8\msun\yr^{-1}$ that happen to be similar to
the rates at which star-formation consumes gas in these galaxies. The accreted gas
is required to have significantly smaller specific angular momentum about the
galaxy's symmetry axis than thin-disc gas.

FB08 suggested that the gas swept up by the fountain's clouds
comes not from the corona but from cool streams embedded within it.  They
were driven to this conclusion by the shortness of the time it takes coronal
gas to flow past a cloud -- this time is very much shorter than the cooling
time of coronal gas.  In this paper we argue that notwithstanding the
disparity between the cooling and flow times of the coronal gas, the
interaction between a cold cloud and coronal gas leads to cooling of the
coronal gas and accretion of it onto the star-forming disc. 

In Section \ref{sec:theory} we give analytical arguments why cloud-corona
interaction must lead to cooling of coronal gas rather than evaporation of \hi.
In Section \ref{sec:simuls} we present hydrodynamical simulations of the flow
past a cloud, which confirm the analytic arguments. In Section
\ref{sec:discuss} we discuss the implications of these simulations both for
further observations and the theory of galaxy formation. Section
\ref{sec:conclude} sums up.

\begin{figure}\label{fig:Ctime}
\centerline{\epsfig{file=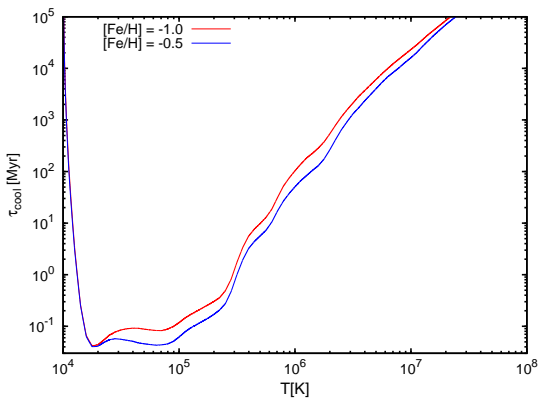,width=.9\hsize}}
\caption{Cooling time using Sutherland \& Dopita (1993) cooling curve for
$\hbox{[Fe/H]}=-1$ and $-0.5$ for material in pressure equilibrium with
coronal gas with $T=1.8\times10^6\K$ and $n_\e=2.6\times10^{-3}\cm^{-3}$.}
\end{figure}

\section{Analytic arguments}\label{sec:theory}

\subsection{Cooling rates}

In Model 2 of \cite{FukugitaP06}, the density of the corona decreases with
radius as $r^{-3/2}$ and at $10\kpc$ has electron density
$n_\e=2.6\times10^{-3}\cm^{-3}$. The corona is assumed to be isothermal with
$T_\h=1.8\times10^6\K$.  \fig{fig:Ctime} shows the cooling time of gas
with metallicities $\hbox{[Fe/H]}=-1$ and $-0.5$ that is in pressure
equilibrium with plasma of this temperature and the mean density of this
corona between 8 and $12\kpc$. At $T\sim2\times10^6\K$, the cooling time is
$\sim300\Myr$.  Once the temperature has fallen to $5\times10^5\K$, the
cooling time has dropped by nearly forty to $\sim8\Myr$, which is less than
the dynamical time at the solar galactocentric radius, $R_0$. By the time the
temperature has dropped by a further factor of two to $2.5\times10^5\K$, the
cooling time has fallen by more than a further order of magnitude and is a
mere $0.4\Myr$. Thus although the cooling time of ambient gas in the lower
corona is long, any diminution in temperature will dramatically shorten the
cooling time. In the model of \cite{FukugitaP06} the radiating plasma flows
inwards at a rate $\sim1\msun\yr^{-1}$. The work done by compression offsets
radiative losses, so the plasma temperature remains $1.8\times10^6\K$.

The mass of an \hi\ halo is rather ill-defined: in an external galaxy it is
not clear at what value of $|z|$ we should place the boundary between disc
and halo \hi, and in the Galaxy there is a similar ambiguity in the
line-of-sight velocity that divides halo from disc \hi.  If in NGC 891 the
disc-halo boundary is conservatively place at $1.3\kpc$, the mass of the \hi\
halo is $\sim6\times10^8\msun$ (FB06). \cite{KalberlaD08} conclude that
$\sim10$ per cent of the Galaxy's \hi\ is halo gas, and that the total \hi\
mass within $12\kpc$ (the region within which most star formation occurs)
is $4.5\times10^9\msun$, so the mass of the \hi\ halo within $12\kpc$ is
$\sim4.5\times10^8\msun$.

The \hi\ halo is largely confined to the region $|z|<5\kpc$, so we compare
the mass of the \hi\ halo at $R<12\kpc$ to the mass of coronal gas in the
cylindrical annulus $4\le R/\!\kpc\le12$, $|z|<5\kpc$.  The coronal mass is
$\sim2\times10^8\msun$, so within this volume there is over twice as much
\hi\ as coronal gas, although the coronal gas will occupy nearly all the
space. 

The clouds that make up the \hi\ halo typically take $100\Myr$ to travel from
their launch point in the plane through the halo and back to the plane.
Consequently, in one Gyr $\sim4.5\times10^9$ of \hi\ is passed through the
$2\times10^8\msun$ of gas in the lower corona. Hence the temperature of the
coronal gas would be halved if just 4.5 per cent of the gas that passes
through the corona in a Gyr were to mix with the coronal gas. This drop in
temperature would bring the cooling time of the coronal gas down to
$\sim 0.1\Gyr$, the local orbital time. 

\subsection{Implications of momentum}

The \hi\ clouds of the halo plough through the corona at speeds $v$ that are less
than but comparable to the sound speed of the coronal gas. Consequently the
flow around these clouds is likely to be in a high Reynolds number regime,
and each cloud must be decelerated by ram pressure of order $\rho_\h v^2$,
where $\rho_\h$ is the coronal density. In these
circumstances the cloud loses its momentum on a timescale equal to
$\rho_\c/\rho_\h$ times the time $t_\c$ for the cloud to move its own length
(FB08). Clouds have diameters of a few tens of parsecs and travel a few kpc,
while $\rho_\c/\rho_\h\simeq200$. Hence clouds must surrender a significant
fraction of their momentum to coronal gas \citep{ben02}. Above we showed that
in a Gyr $\sim4.5\times10^9\msun$ of \hi\ passes through the
$2\times10^8\msun $ of the lower corona. Clearly the coronal gas cannot
absorb a significant fraction of the momentum of more than 20 times its mass
of \hi\ (FB08). Moreover, if the \hi\ disc were losing angular momentum to
the corona at the rate this calculation implies, it would be contracting on a
Gyr timescale. The natural resolution of these problems is that the disc {\it
accretes\/} most of the coronal gas that \hi\ clouds encounter. Then the
momentum lost by \hi\ clouds would be returned to the disc, and it would not
build up in the corona.

\subsection{Relation to cooling flows}

As was mentioned in Section 1, coronal gas in dark-matter halos more massive
than those of spiral galaxies has been extensively studied for four decades.
Cooling within these halos shows no tendency to produce a cold stellar disc
-- the coldest gas is near the centre, where the cooling time is shortest,
and whatever gas cools out of the corona feeds the central black hole rather
than forming a star-forming disc.  The argument of the last subsection may
help us to understand why in less massive dark halos cooling coronal gas
flows into the disc rather than onto a central black hole: so long as the
halo has a star-forming disc, that disc sustains its star formation by
reaching up and grabbing coronal gas.

Discs are disrupted by major mergers, which occur rather frequently.
Analytic arguments and hydrodynamical simulations of cosmological clustering
suggest that when the disc of a relatively low-mass galaxy is disrupted by a
major merger, it will quickly re-form from filaments of cold inflowing gas
\citep[e.g.][]{Governato07}.  Thereafter it will sustain itself by grabbing
coronal gas.  When a more massive galaxy experiences a major merger, the
disrupted star-forming disc is less likely to re-form. If it does not form
from cold inflowing gas, it will not form from coronal gas, because in the
absence of a star-forming disc, catastrophically cooling coronal gas will
feed the central black hole and reheat the corona. Thus star-formation
permanently ceases if there is insufficient cool gas to re-form a
star-forming disc after a merger. 

In this picture it is likely that cooling at the centre of the corona of a
disc galaxy leads to episodic reheating of the central corona, just as in a
classical cooling flow. That is, we suggest that the corona of a star-forming
disc galaxy accretes onto central black hole {\it as well as\/} onto the
disc. For this to be a viable proposal, the central reheating associated with
accretion by the black hole must not undermine the ability of the
star-forming disc to grab coronal gas from the part of the corona that lies
above it. This condition could be satisfied if the AGN outburst were
sufficiently small and sufficiently directed perpendicular to the galactic
plane. This is a topic for a later paper, however.

\subsection{The wake of a typical cloud}

As a cloud moves through the ambient coronal gas, turbulence in the boundary
layer at the interface of the hot and cold fluids must cause gas to be
stripped from the cloud at some rate. To calculate this rate from ab-initio
physics is a daunting task because one would have to consider plasma
instabilities in addition to hydrodynamical ones such as the Kelvin-Helmholtz
instability, and turbulent energy will be cascading to very small scales. In
view of these difficulties, the obvious way forward is to parametrise the
problem by hypothesising that mass is stripped from the a cloud of mass $M_\c$
at a rate $\alpha M_\c$.

A turbulent wake of stripped gas will run back through the corona from the
moving cloud. In this wake turbulence will mix the stripped gas with
ambient gas. A key quantity is the cooling time of the plasma
that results from this mixing. We estimate this under the assumption that
mixing occurs so quickly that radiative losses during mixing can be
neglected.

Let $s$ be a short length of the wake, which has cross-sectional area
$A_\w$ and originally contained a mass $A_\w s\rho_\h$ of coronal gas.
In the time $s/v$ that it took the cloud to pass through the length, a mass
$\alpha M_\c s/v$ of gas was stripped from it. After the cold gas has mixed
and come into thermal equilibrium with the coronal gas, the temperature of
the resulting fluid is
 \begin{equation}
T_\m={A_\w \rho_\h T_\h+\alpha M_\c T_\c /v\over
A_\w \rho_\h +\alpha M_\c /v},
\end{equation}
 where $T_\h$ and $T_\c$ are the coronal and cloud  temperatures,
respectively.

The natural unit for the cross-sectional area of the wake is the
characteristic cross-section $(M_\c/\rho_\c)^{2/3}$ of the cloud. We write
$A_\w= \beta(M_\c/\rho_\c)^{2/3}$, where $\beta\sim1$. Thus
 \begin{equation}
T_\m={\beta(M_\c/\rho_\c)^{2/3} \rho_\h T_\h+\alpha M_\c T_\c /v\over
\beta(M_\c/\rho_\c)^{2/3} \rho_\h +\alpha M_\c /v}.
\end{equation}
 We simplify this expression under the assumption that the cloud is in approximate pressure equilibrium with the
corona, so $\rho_\h T_\h\simeq\rho_\c T_\c$, and have
 \begin{equation}\label{eq:Trat}
T_\m={1+ (M_\c/\rho_\c)^{1/3}\alpha/\beta v\over
1 + (M_\c/\rho_\c)^{1/3}(T_\h/T_\c)\alpha /\beta v}\,T_\h.
\end{equation}
 Since $(M_\c/\rho_\c)^{1/3}\alpha/v$ is the fraction of the cloud's mass
that is stripped in the time taken for the cloud to travel its own length,
and $\beta\sim1$, the second term in the numerator of equation
(\ref{eq:Trat}) must be small compared to unity and may be neglected. The
second term in the denominator is larger by a factor $T_\h/T_\c\ga200$, so
$T_\m$ may be significantly lower than $T_\h$. On account of the steepness of
the cooling-time curve plotted in \figref{fig:Ctime}, this result implies that
the cooling time in the wake may fall below the ambient cooling time by a
factor of several.

From this back-of-envelope calculation we draw the following conclusions

\begin{itemize}
\item Material in the wake will become \hi\ on the timescale that it takes
the parent \hi\ cloud to fly its trajectory if the mass-loss rate $\alpha$
exceeds the critical value
 \begin{equation}\label{eq:defsac}
\acrit\equiv{\beta v\over(M_\c/\rho_\c)^{1/3}}\,{T_\c\over T_\h}.
\end{equation}
 For $\beta=1$ this
condition is that in the time taken to travel its own length the cloud lose
at least a fraction $T_\c/T_\h$ of its mass.

\item If the mass-loss rate of clouds falls below this critical value, what
mass is stripped from the cloud will be integrated into the corona. This will
lower the cooling time of the ambient corona, but not lead to prompt
accretion of the wake onto the disc. This drop in the temperature of the
corona near the disc will lower the critical mass-loss rate required for
subsequent wakes to cool promptly.

\item Our estimate of $\acrit$ was obtained
under the assumption that we can neglect  cooling during mixing. Although the
mixing process is likely to be fast, the cooling rate is extremely large at
temperatures that lie within a factor 30 of the cloud's temperature. Hence it
is likely that a more exact calculation would produce a lower  estimate of
$\acrit$. We have also neglected compression of the coronal gas as
it flows around the cloud. However, the effect of  compression on the cooling
time of ambient gas is unclear because, while the cooling time decreases with
increasing density at constant temperature, it increases with temperature, and
compression will be associated with a (largely adiabatic) rise in $T$.

\item The effective value of $\beta$ is uncertain. On one hand, as the
simulations will show,  the cloud flattens in its direction of motion, making
$\beta>1$. On the other hand, the distribution of cloud material is likely to
be concentrated in a network of thin sheets. The effective value of $\beta$
for an individual sheet could be small.

\item Dimensional analysis indicates that $\alpha$ will lie near $\acrit$: the rate of stripping is essentially determined by the rate at which
coronal gas hits the leading surface of the cloud and strips  a comparable
mass from the cloud. Quantitatively, $\alpha M_\c\equiv\dot M_\c=-b\pi r^2\rho_{\rm h}v$,
where $b\sim1$. When we eliminate $r$ and $\rho_\h$ in favour of $M_\c$ and
$T_\h$ we find that the characteristic mass-loss parameter is
 \begin{equation}
\alpha=\left({9\pi\over16}\right)^{1/3}{b\over\beta}\,\acrit.
\end{equation}

\item If we take the mass-loss rate to be given by the previous item, we find
that the mass of a cloud that will completely mix with the corona after
travelling distance $L$ is
\begin{equation}\label{eq:MofL}
M_{\rm crit}={9\pi\over16}\left({T_\c\over T_\h}bL\right)^3\rho_\c.
\end{equation}
 Equivalently, the distance travelled prior to destruction is predicted to be
$\sim T_\h/ T_\c$ times the size of the cloud.

\end{itemize} 

\section{numerical simulations}\label{sec:simuls}

Numerical simulations of a cloud of gas at $T\sim10^4\K$ moving through
coronal gas at an initial speed $v_0\sim75\kms$ will illustrate these points
and give insight into both typical mass-loss rates and the critical mass-loss
rate required for prompt cooling of the wake. The simulations are idealised
in that they neglect gravitational acceleration and the variation 
of the coronal density along the cloud's trajectory. However, the galactocentric radius of the
clouds varies by $\la30$ per cent, so the density variation is not extreme.
The range of relevant velocities depends on the extent to which the corona
corotates with the disc, which is currently unclear. The velocity we have
chosen to explore is at the low end of the relevant range because it is of
order the velocity at which clouds have to be ejected from the disc in order
to reach heights of a few kiloparsecs. Larger velocities would clearly lead
to more rapidly interaction with the corona than that explored here.

It is important to be clear about what can and cannot be learnt from the
simulations. First any simulation incorporates limited physics and limited
spatial resolution. In the real world the fluid is a magnetised, almost
perfectly conducting, largely collisionless plasma. The shear flow at the
cloud-corona interface will draw out and strengthen the field lines
originally in the plasma. This drawing out will make the velocity
distribution of particles anisotropic, which will excite plasma
instabilities. These instabilities will both heat and mix the plasma. By
contrast the tendency of the field lines to follow stream lines will strongly
inhibit mixing. None of this complex physics is included in our simulations.

Instead, in the simulations ablation and mixing are driven by the
Kelvin-Helmholtz instability, which gives rise to ripples in the fluid
interface. These ripples develop into vortices, which shed smaller vortices,
which themselves shed vortices.  In a numerical simulation numerical
viscosity prematurely truncates this hierarchy of vortices on a scale of a
few times the grid spacing $\Delta$.  On scales finer than $\Delta$, the
fluid is represented as perfectly mixed, whereas it is in reality a roughly
fractal foam of high- and low-density regions. The condition for the
simulations to be a reliable guide to the large-scale structure of the flow
is that the large-scale dynamics of this foam is equivalent to the dynamics
of the locally homogeneous fluid that is actually represented. This is a
reasonable proposition, but we have to expect that the values of the flow's
macroscopic parameters, such as ambient pressure and cooling rate, at which a
simulation of given resolution most closely approximates reality are likely to
vary with resolution. That is, we anticipate a phenomenon analogous to
renormalisation in quantum field theory, where the appropriate value of the
bare electron mass, for example, is a function of the largest wavenumber
summed over. 

The simulations do not include thermal conduction. On the smallest scales
conduction must play an important role in homogenising a mixture of cloud and
coronal gas, and in the simulations this role is effectively covered by
numerical mixing and diffusion. Consequently, our failure to model thermal
conduction is most worrying on intermediate and larger scales. In a
magnetised plasma heat is largely conducted along field lines. The field
lines in upstream ambient gas are inevitably disconnected from field lines in
the cloud. Any connection between these sets of field lines must occur in the
turbulent wake. This fact is likely to severely limit the effectiveness of
thermal conduction. 

The physical properties of the clouds and the corona considered
here are such that the clouds would be stable against
conductively-driven evaporation if they were stationary
\cite[see][]{NipotiB07}, so it is very unlikely that conduction
plays an important role in the energetics of our problem. In
principle conduction lowers the rate at which a moving cloud is ablated, by
damping the Kelvin-Helmholtz instability \citep{Vieser07}, but this effect
will be unimportant if conduction is magnetically suppressed to values small
compared to the Spitzer or saturated values.

In light of these remarks,
the aims of the simulations are as follows

\begin{itemize}
\item To estimate the mass-loss rate $\alpha$ for comparison with
$\alpha_{\rm crit}$. It is advantageous to do this
in the absence of radiative cooling for then (a) the calculations are slightly faster,
and (b) one can identify the cloud with gas at $T\la10^5\K$ since any ablated
material will be heated to and remain at higher temperatures. We will show
that the mass-loss rate is reasonably independent of $\Delta$.

\item To show that for any given metallicity of the gas, there is a critical
ambient pressure $P_{\rm crit}$ above which the mass of cool gas increases
with time through condensation in the wake and below which the wake tends to
evaporate. We shall find that although our values of $P_{\rm crit}$ vary with
both metallicity and $\Delta$, they lie within the range of values that occur
in practical cases. Hence it is plausible that the true value of $P_{\rm
crit}$ lies below the actual ambient pressures, so real wakes give rise to
condensation and accretion.

\end{itemize}

\begin{table*}
\caption{Parameters of the simulations. The grids sizes are coarse (c)
$384\times768$, medium (m) $512\times1024$ or $1536$ depending  on
duration, and fine (f) $768\times1536$. Each configuration is simulated both
with and without radiative cooling.}
\label{tab:simuls}
\begin{tabular}{|llllllllr|}
\hline
simulation & time [Myr] & \feh  & $T_{\rm h} [\!\K]$ &
$n_{\rm h} [\!\cm^{-3}]$ &
$T_{\rm c} [K]$& $M_\c [10^4\msun]$ & grid size \\
\hline
\multicolumn{8}{|c|}{Low-pressure simulations} \\
\hline
Z\_0 & 25 & no met.  & $2\times 10^6$ & $10^{-3}$ & $10^4$ & $1.1$ & c,m,f \\
Z\_1 & 25 & -3.0  & $2\times 10^6$ & $10^{-3}$ & $10^4$ & $1.1$ & c,m,f \\
Z\_2 & 25 & -2.0  & $2\times 10^6$ & $10^{-3}$ & $10^4$ & $1.1$ & c,m,f \\
Z\_3 & 25 & -1.5  & $2\times 10^6$ & $10^{-3}$ & $10^4$ & $1.1$ & c,m,f \\
Z\_4 & 25 & -1.0  & $2\times 10^6$ & $10^{-3}$ & $10^4$ & $1.1$ & c,m,f \\
Z\_5 & 25 & -0.5  & $2\times 10^6$ & $10^{-3}$ & $10^4$ & $1.1$ & m,f \\
Z\_6 & 25 &  ~0.0  & $2\times 10^6$ & $4 \times 10^{-4}$ & $10^{4}$ & $0.44$ & m \\
\hline
\multicolumn{8}{|c|}{High-pressure simulations} \\
\hline
T\_4\_Z\_3 & 25 & -1.5 & $2\times 10^6$ &
$2\times10^{-3}$ & $10^4$ & $2.2$ & m\\
T\_4\_Z\_4 & 50 & -1.0  & $2\times 10^6$ & $2\times10^{-3}$ &
$10^4$ & $2.2$ & m\\
T\_3\_Z\_4 & 50 & -1.0  & $2\times 10^6$ & $2\times10^{-3}$ &
$5\times 10^3$ & $4.4$ & m\\
\hline

\end{tabular}
\end{table*}
\subsection{The simulations}

The parameters of the simulations are listed in Table~\ref{tab:simuls}.  In
all simulations the initial cloud velocity was $75\kms$ and 
the initial radius was $100\pc$. \rev{The temperature of the corona is
restricted to a narrow range around $2\times10^6\K$ by the requirement that
the corona be bound to the Galaxy and yet be extensive enough to contain a
cosmologically significant mass \citep[e.g.]{FukugitaP06}, so in all
simulations we made the corona's temperature $2\times10^6\K$.  In simulations
of the low-pressure sequence the total particle density of the corona was
$10^{-3}\cm^{-3}$ (except in the Z\_6 simulation in which it was reduced to $4 \times 10^{-4}\cm^{-3}$),
implying $n_\e\simeq0.5\times10^{-3}$, while in the
high-pressure simulations it was twice as great. These values are both lower
than the density $n_\e=2.6\times10^{-3}\cm^{-3}$ at $r=10\kpc$ in model 2 of
\cite{FukugitaP06}, and may be compared with the total particle density
$n=4\times10^{-4}$ at $10\kpc$ above the plane adopted by \cite{HeitschP09}.
In all but two simulations the initial cloud temperature was $10^4\K$; in
the last two (high-pressure) simulations the cloud temperature was lowered
to $5\times10^3\K$ since at the higher temperature (and therefore lower
density contrast) the cloud was totally disrupted by $50\Myr$.} The cloud mass ranged
from $0.44$ to $4.4\times10^4\msun$ depending on the pressure of the corona
and the temperature of the cloud. The Jeans mass of the standard cloud
($r = 100$ pc, $T_{\rm c} = 10^4$ K) is
$1.6\times10^8\msun$ so our neglect of self gravity is amply justified.

The calculations were performed on
two-dimensional, Cartesian grids of three sizes: $384\times768$ (c),
$512\times1024$ or $1536$ for simulations run to $50\Myr$ (m), and
$768\times1536$ (f). Ghost cells are used to ensure that the pressure
gradient vanishes at the grid boundaries.  Every configuration was simulated
twice, once with cooling on and once with cooling off. When radiative cooling is
permitted, it follows the prescription of \cite{SutherlandD}.  The metallicity
of the cloud is always the same as that of the ambient medium, and is varied
from zero up to solar.
We used the Eulerian code ECHO, which is
flux-conserving and uses high-order shock-capturing schemes; a detailed
description and tests of it can be found in \cite{Londrillo}. 
 
Since one of the dimensions perpendicular to the cloud's velocity has been
suppressed, we are in effect simulating flow around an infinite cylindrical
cloud that is moving perpendicular to its long axis. The cylinder initially
has a circular cross section of radius $r$. From the simulations we obtain
quantities per unit length of the cylinder. We relate these to the
corresponding quantities for an initially spherical cloud of radius $r$ by
multiplying the cylindrical results by the length ${4\over3}r$ within which
the mass of the cylinder equals the mass of the spherical cloud.

\begin{figure}
\centerline{\epsfig{file=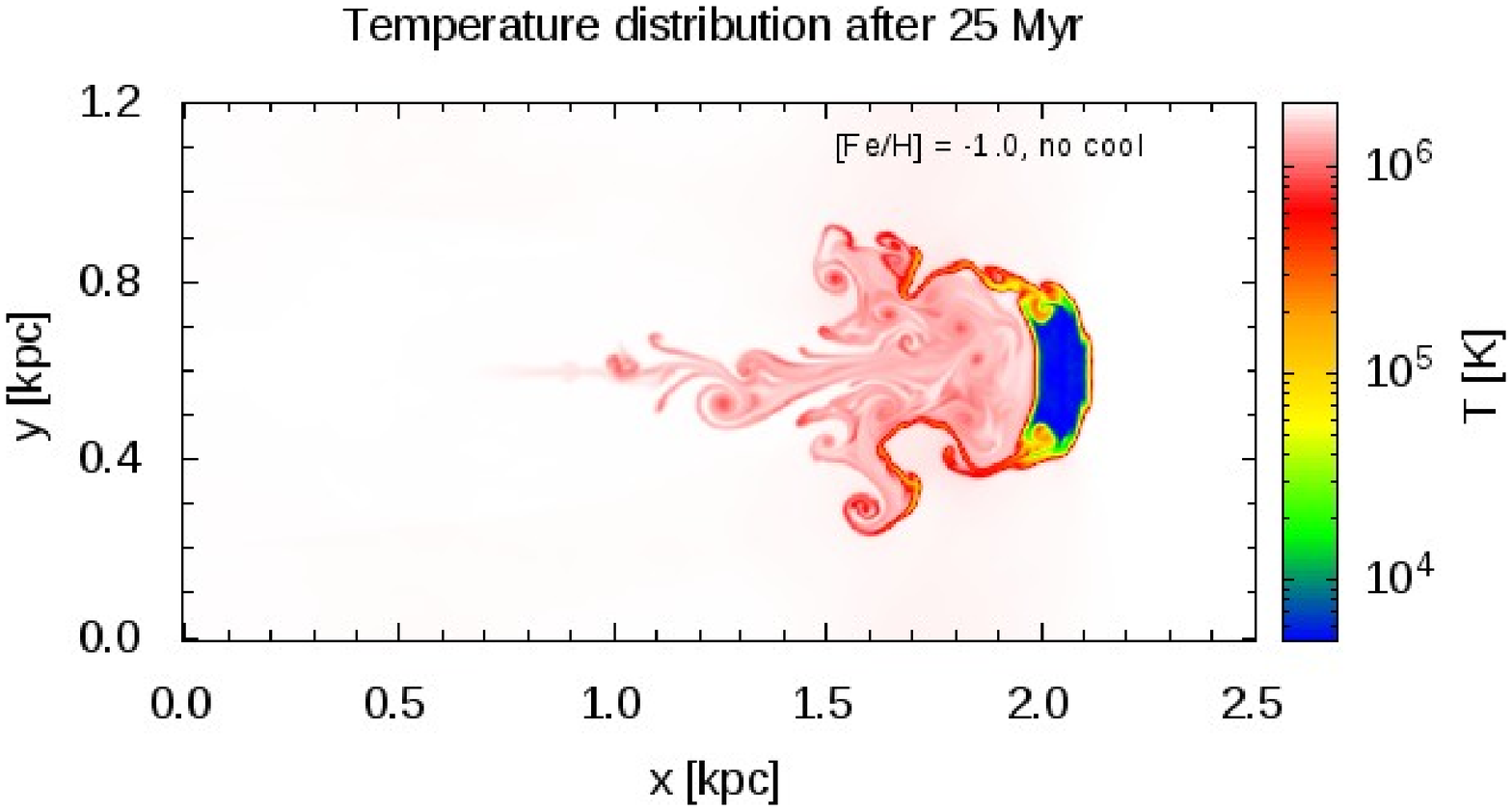,width=\hsize}}
\centerline{\epsfig{file=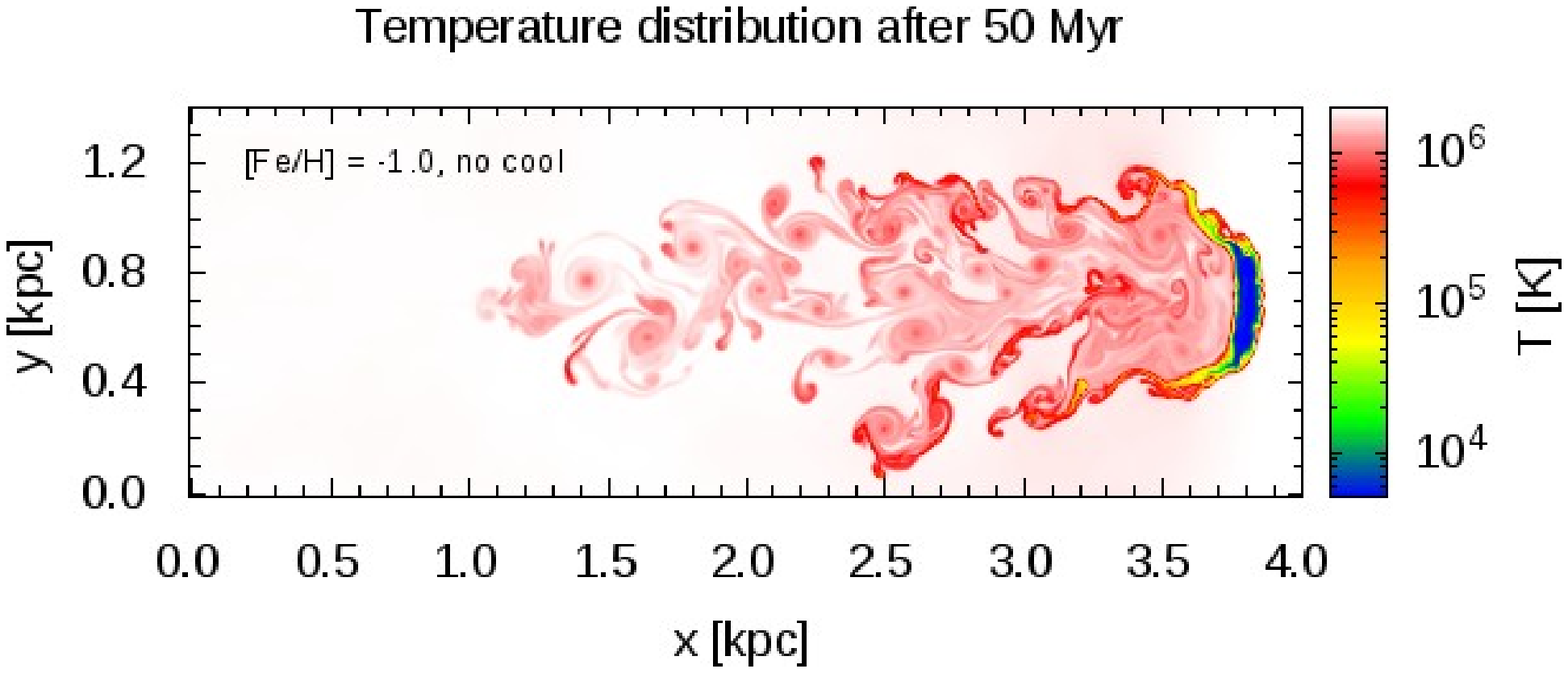,width=\hsize}}
\caption{The temperature distribution after $25$ and $50\Myr$ when radiative cooling is switched
off. The computational grid has medium resolution (see table \ref{tab:simuls}), the initial cloud
temperature was $5\times10^3\K$ and the coronal density was
$2\times10^{-3}\cm^{-3}$.}\label{fig:Tplot}
\end{figure}

\subsection{Flows without radiative cooling}

We first study the simulated flows in the absence of radiative cooling, in
order to assess the importance of the limited resolution of the simulations.

\fig{fig:Tplot} shows the temperature distribution on the grid after
$25\Myr$ and $50\Myr$ with radiative cooling turned off. The cloud has been
flattened into a pancake by ram pressure from the medium it is moving into.
The shear flow over the leading face of the pancake is causing vortices to be
shed from the pancake's edges that are analogous to the vortices shed by an
aeroplane wing.  In the highly turbulent wake behind the cloud, the temperature
fluctuates around $10^6\K$ depending on the fraction of the gas in each cell
that comes from the cloud rather than the corona. 

\begin{figure}
\centerline{\epsfig{file=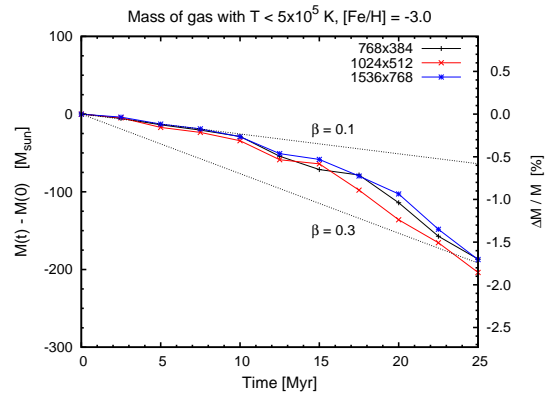,width=.9\hsize}}
\caption{The evolution of the mass of gas at $T<5\times10^5$ for three
different resolutions when radiative cooling is switched off. The dotted
straight lines
show the effect of the critical mass-loss rate $\alpha_{\rm crit}$ for two
values of $\beta$. The
particle density of the corona is $10^{-3}\cm^{-3}$.}
\label{fig:nocoolablate}
\end{figure}

\fig{fig:nocoolablate} quantifies the speed at which ablation reduces the
cloud's mass by plotting the mass of gas below $5\times10^5\K$ versus time
for simulations of three resolutions. We see that over $25\Myr$
$\sim200\msun$ of gas ($\sim2$ per cent of the cloud's mass) are heated to
above $5\times10^5\K$, regardless of the grid resolution; it seems that even
the coarsest grid has sufficient resolution to model satisfactorily the
stripping of gas from the leading edge of the cloud. In the absence of
radiative cooling, any gas that is stripped from the cloud will eventually be
heated to above $5\times10^5\K$ as it mixes with coronal gas. The details of
the smallest vortices involved in the mixing process are resolution-dependent
in the sense that at higher resolutions some gas remains cold for slightly
longer before numerical mixing on the grid scale eliminates it. The cloud's
mass-loss rate is resolution-independent because all stripped gas will be
heated within a couple of large-scale eddy-turnover times, regardless of
resolution.

The mass-loss rate increases gently with time because it depends on the area
of the cloud's leading face, which increases with time as the cloud is
squashed into a thinner and thinner pancake, a process that is apparent in
\figref{fig:Tplot}.

The dotted straight lines in \figref{fig:nocoolablate} show the critical
mass-loss rate defined by equation (\ref{eq:defsac}) for $\beta=0.1$ and
$0.3$. We see that, as predicted above on dimensional grounds, the measured
mass-loss rate lies near $\acrit$.  Actually the simulations must
underestimate $\dot M_\c$ because mass is lost from the cloud's edges, which
have total length ${8\over3}r$ in cylindrical geometry and $2\pi r$ in the
spherical case. Thus we expect the numerical rates to be $\sim0.42$ of the
true rate and in three dimensions $\beta$ would lie close to unity.

These results enable us to estimate the minimum size that a cloud must have
if it is to survive a typical passage through the corona. From Fig.~10 of
FB06 we have that trajectories last $100\Myr$, and in this
time a cloud travelling at $v\sim70\kms$ will move $7\kpc$. Setting
$L=7\kpc$ and $b=1$ in equation (\ref{eq:MofL}) we find that
clouds with masses less than 
\begin{equation}\label{eq:mcrit}
M_{\rm
crit}=220\left({L\over7\kpc}\right)^3\left({T_\h/T_\c\over200}\right)^{-2}
\left({n_\h\over10^{-3}\cm^{-3}}\right)\msun
\end{equation}
 will completely mix with the
corona before they can return to the plane.

It is important to know how the mass-loss rate scales with cloud mass and
thus cloud radius. Since mass is lost from the leading surface of the cloud,
we would expect the absolute mass-loss rate to scale as $M_\c^{2/3}$ and
therefore the specific mass-loss rate $\alpha\propto M_\c^{-1/3}$. We have
verified this dependence by  simulating the evolution of a cloud of
the standard density but radius doubled to $200\pc$.

\begin{figure}
\centerline{\psfig{file=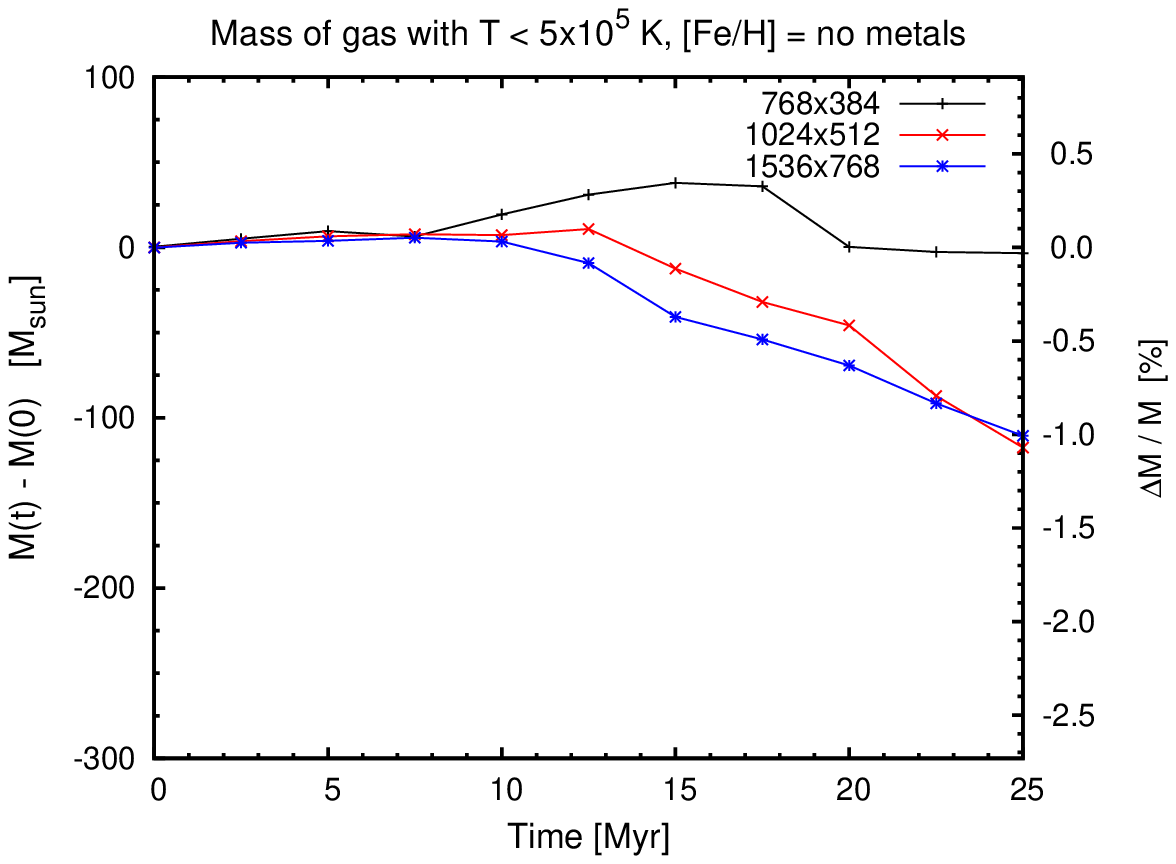,width=.9\hsize}}
\centerline{\psfig{file=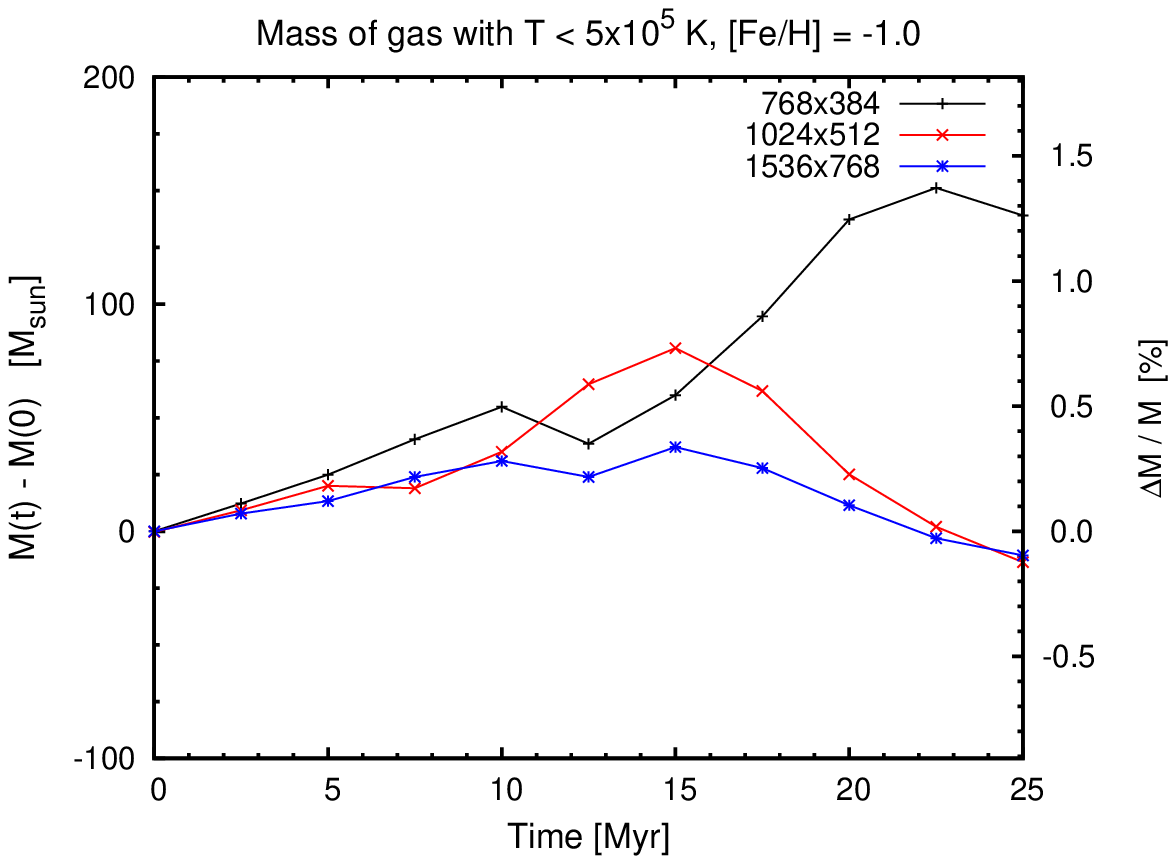,width=.9\hsize}}
\centerline{\psfig{file=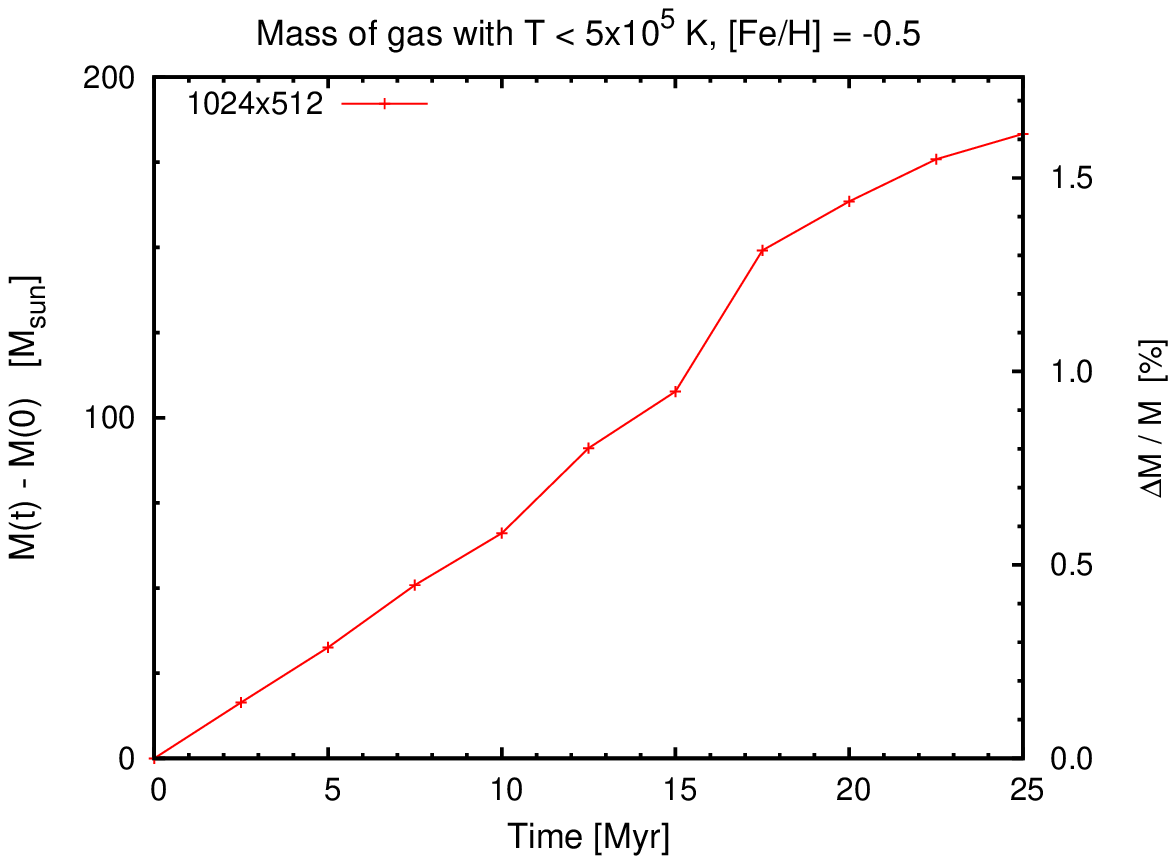,width=.9\hsize}}
\caption{The same as \figref{fig:nocoolablate} but with radiative cooling turned on.
In the top panel the plasma has primordial abundances, while in the middle
and bottom
panels $\feh=-1$ and $-0.5$, respectively. In all three panels the corona had
particle density $10^{-3}\cm^{-3}$.}\label{fig:cool-0}
\end{figure}

\subsection{Flows with radiative cooling}

When radiative cooling is switched on, the strength of the dependence of the
cooling rate upon $T$ that is apparent in \figref{fig:Ctime}, substantially
increases the difficulty of the simulations and the uncertainties surrounding
their results because now the structure of the turbulent wake is crucial.
Indeed,  evaporation is favoured over cooling by more effective dispersal of
stripped material through a large volume of the corona. The higher the
resolution delivered by the code, the greater the dynamic range of the
hierarchy of vortices, and the more effective is the dispersal of stripped
material, with the consequence that increased resolution favours evaporation
over condensation.

\begin{figure}
\centerline{\epsfig{file=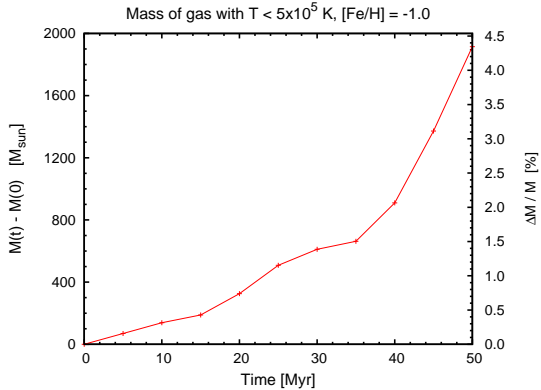,width=.9\hsize}}
\caption{The mass of gas at $T<5\times10^5\K$ in a simulation with coronal
particle density $2\times10^{-3}\cm^{-3}$ and metallicity
$\feh=-1$.}
\label{fig:cool-hp}
\end{figure}

Each panel of \figref{fig:cool-0} is the analogue of
\figref{fig:nocoolablate} but with radiative cooling turned on; in the three
panels the metallicity increases from zero (top panel) through $\feh=-1$
(middle panel) to $-0.5$ (bottom panel). 
\rev{The simulations with [Fe/H] $<$ -1 do not differ significantly from  
that at zero metallicity.} In all three panels the mass of
cool gas starts by increasing rather than decreasing, and the rate of
increase naturally increases with metallicity. When $\feh=-0.5$,
the increase continues throughout the $25\Myr$ simulated, but at lower
metallicities the mass of cool gas eventually starts to decrease.  As
anticipated, the results are unfortunately dependent on numerical resolution:
in general there is a tendency for the mass of cold gas to decrease as the
resolution increases, although the medium and high-resolution simulations
give rather similar results.

\fig{fig:cool-hp} shows the effect of doubling the particle density of the
corona to $2\times10^{-3}\cm^{-3}$: even at metallicity $\feh=-1$ the mass of
cool gas now increases throughout the $50\Myr$ simulated, and in fact in the
latter half of the simulation cold gas accumulates at an accelerating rate.
This result should be contrasted with that shown by the middle panel of
\figref{fig:cool-0}, which shows that at the same metallicity but half the
density, the mass of cool gas starts to decrease after $\sim15\Myr$.
If we extrapolate the behaviour shown in \figref{fig:cool-hp} to the whole
Milky Way halo, assuming as its total mass the value
given in Sect. 2.1, we obtain a global accretion rate of $\approx 0.5\msunyr$.

\begin{figure}
\centerline{\psfig{file=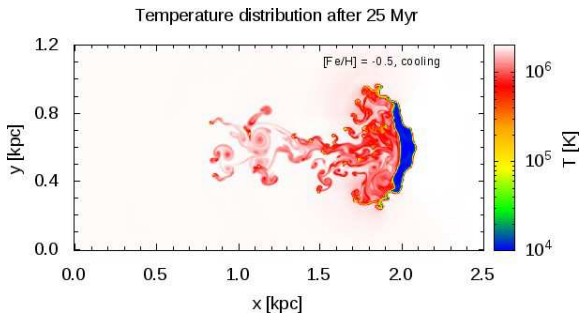,width=.9\hsize}}
\caption{As \figref{fig:Tplot} but with the coronal density lowered to
$10^{-3}\cm^{-3}$ and radiative cooling turned on. The metallicity is
$\feh=-0.5$.}\label{fig:coolTplot}
\end{figure}

\fig{fig:coolTplot} shows the temperature distribution at $t=25\Myr$ in a
simulation with cooling of plasma with $\feh=-0.5$ and ambient particle
density $10^{-3}\cm^{-3}$. Comparing this figure with the upper panel of
\figref{fig:Tplot} we see that cooling makes the wake longer and less
laterally extended.
\rev{We have also run simulations with lower values for the coronal gas
density and found that accretion is still present for 
$n_{\rm h} \gsim 4 \times 10^{-4}\cm^{-3}$ provided that the metallicity is
about solar.
}

\begin{figure}
\centerline{\epsfig{file=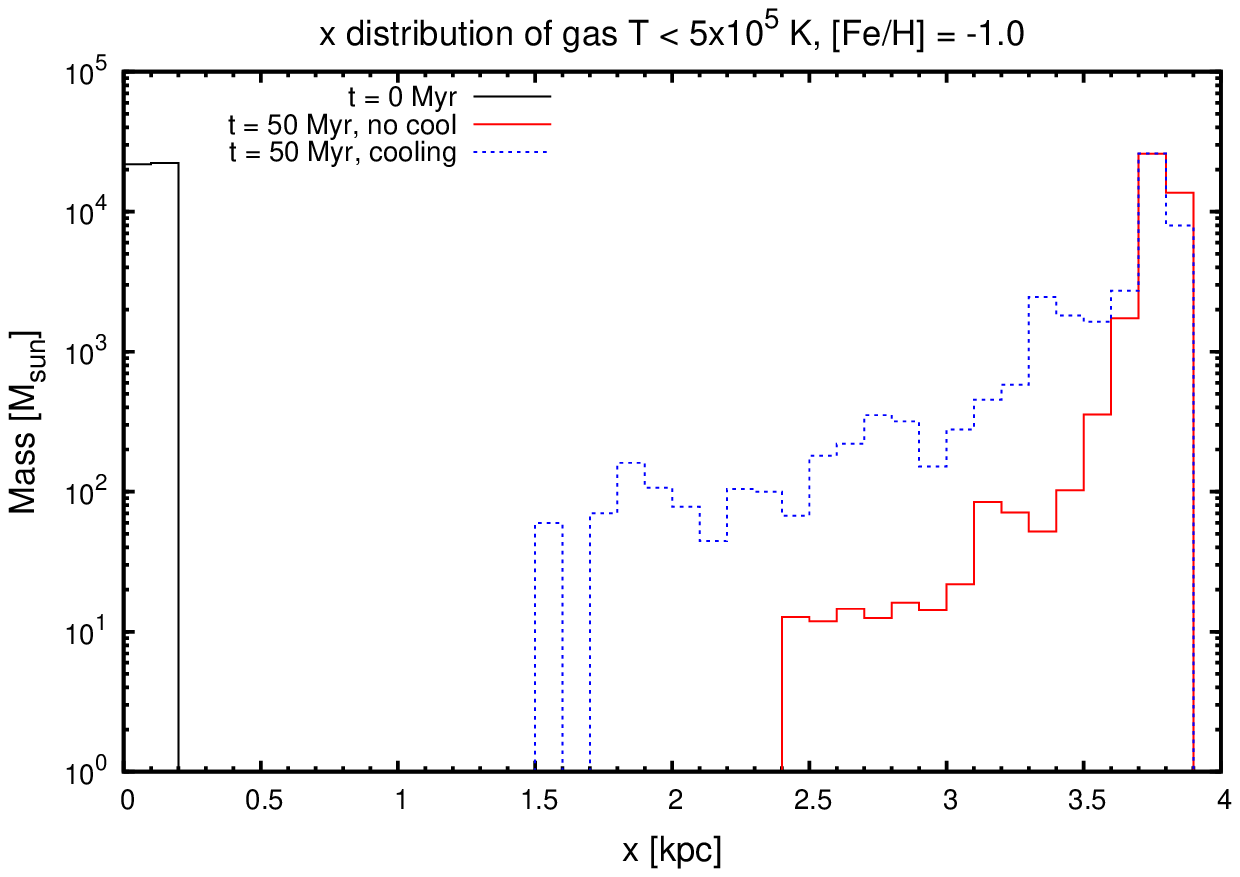,width=.9\hsize}}
\centerline{\epsfig{file=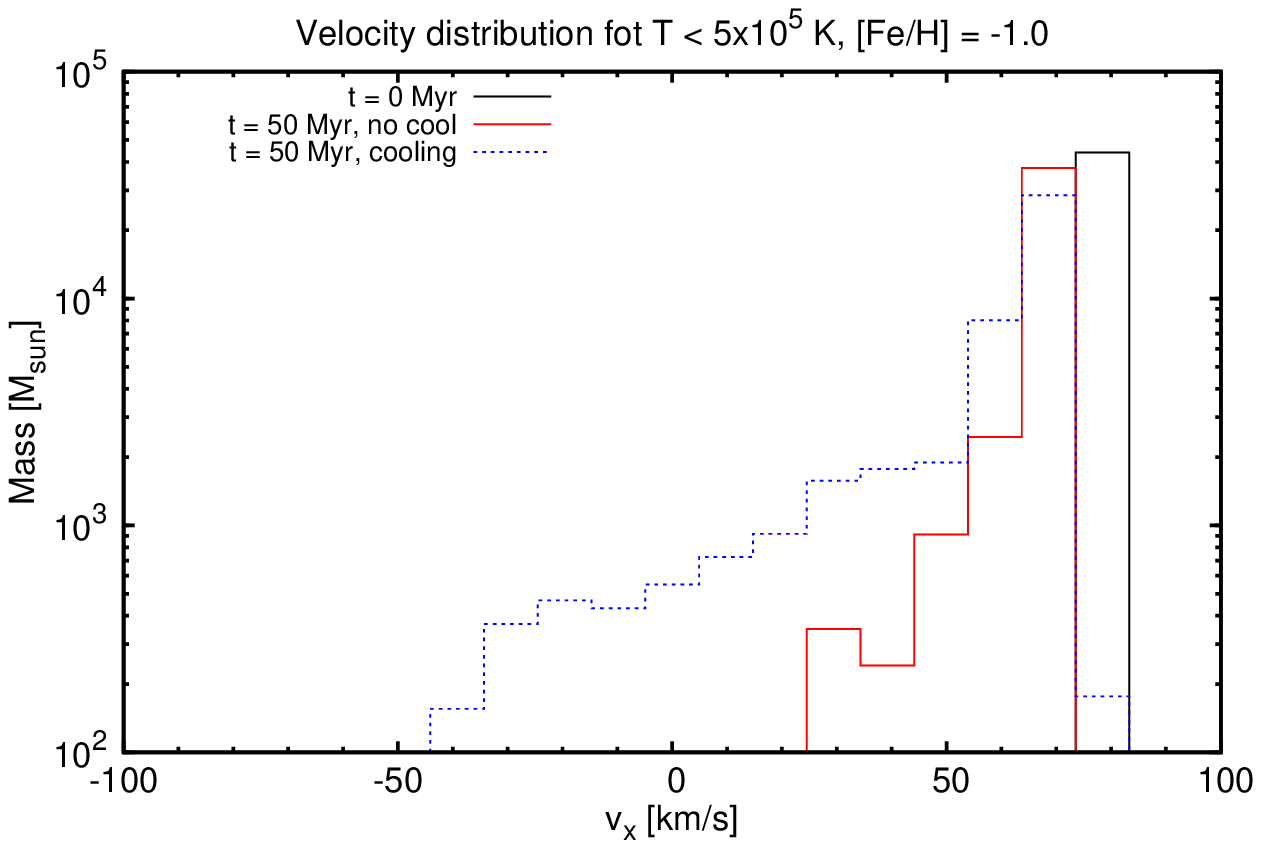,width=.9\hsize}}
\caption{Upper panel: the distribution of cool gas
along the cloud's direction of travel. The black curve shows the initial
distribution while the blue dotted curve and red solid curve show the distributions  at
$t=50\Myr$ with and without radiative cooling, respectively. Lower panel: the
same but for the velocity along the direction of travel.}\label{fig:x-vx}
\end{figure}

\fig{fig:x-vx} shows for the high-pressure simulation T\_3\_Z\_4 the
distribution of cool gas along the cloud's direction of travel (top panel)
and its distribution in velocity along the same direction (bottom panel) at the start
of the simulation (black curve) and after $50\Myr$ when radiative cooling is
(blue curve) or is not (red curve) included. From the upper panel we
clearly see the effectiveness of cooling in enhancing the mass of cool gas.
We also see that $\la1$ percent of the cold gas lies more that $1\kpc$ behind
the cloud. In the lower panel around $75\kms$ we clearly detect the
deceleration of the main body of the cloud, but more striking is the width of
the velocity range over which small amounts of cool gas are distributed. A
few times $100\msun$ is accelerated to higher velocities than the cloud's. A
slightly larger mass of gas is decelerated to negative velocities. This
velocity distribution implies that gas circulates around vorticies at speeds
that are comparable to the speed of the cloud's forward motion.

\begin{figure}
\centerline{\epsfig{file=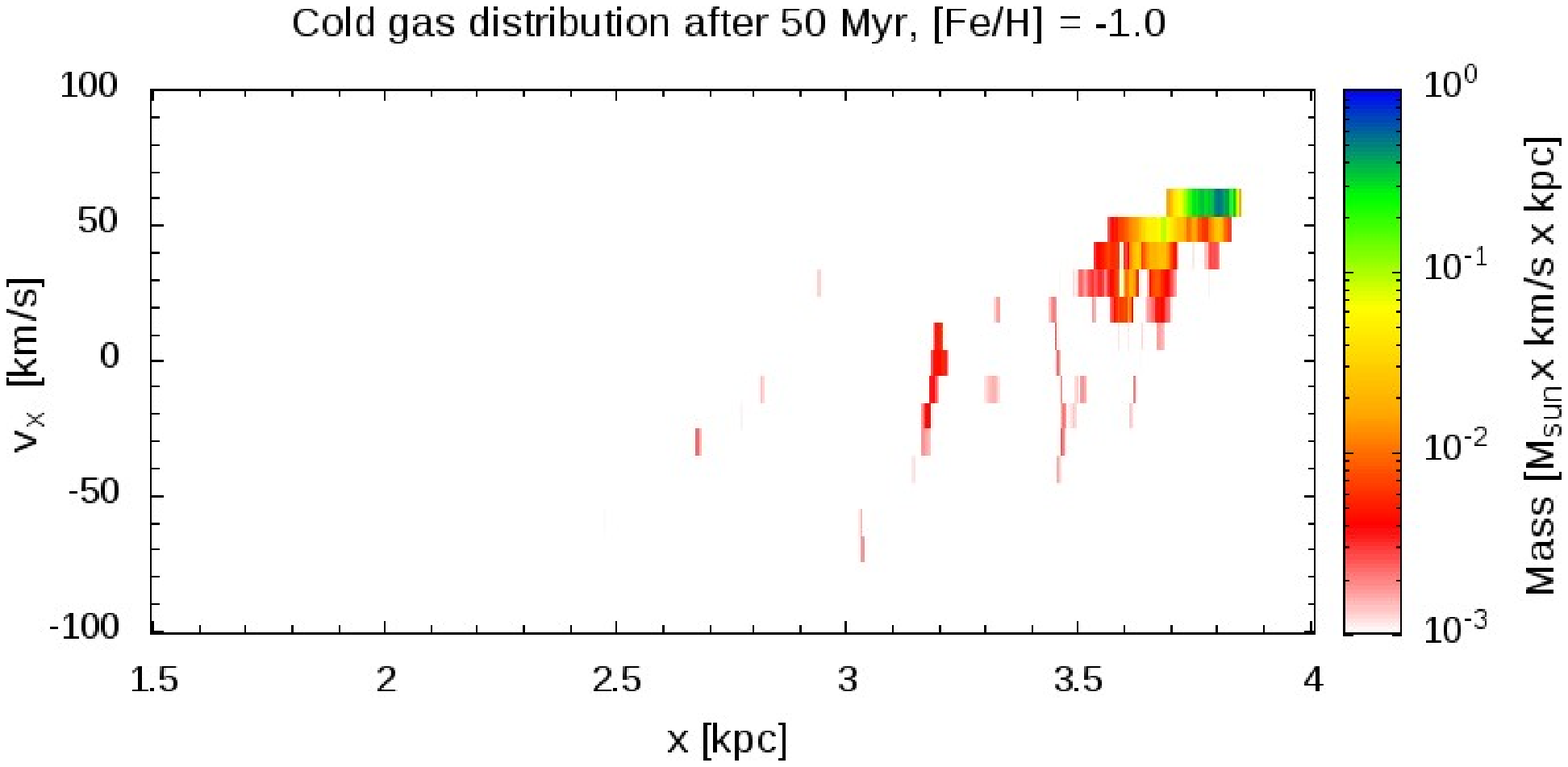,width=.9\hsize}}
\centerline{\epsfig{file=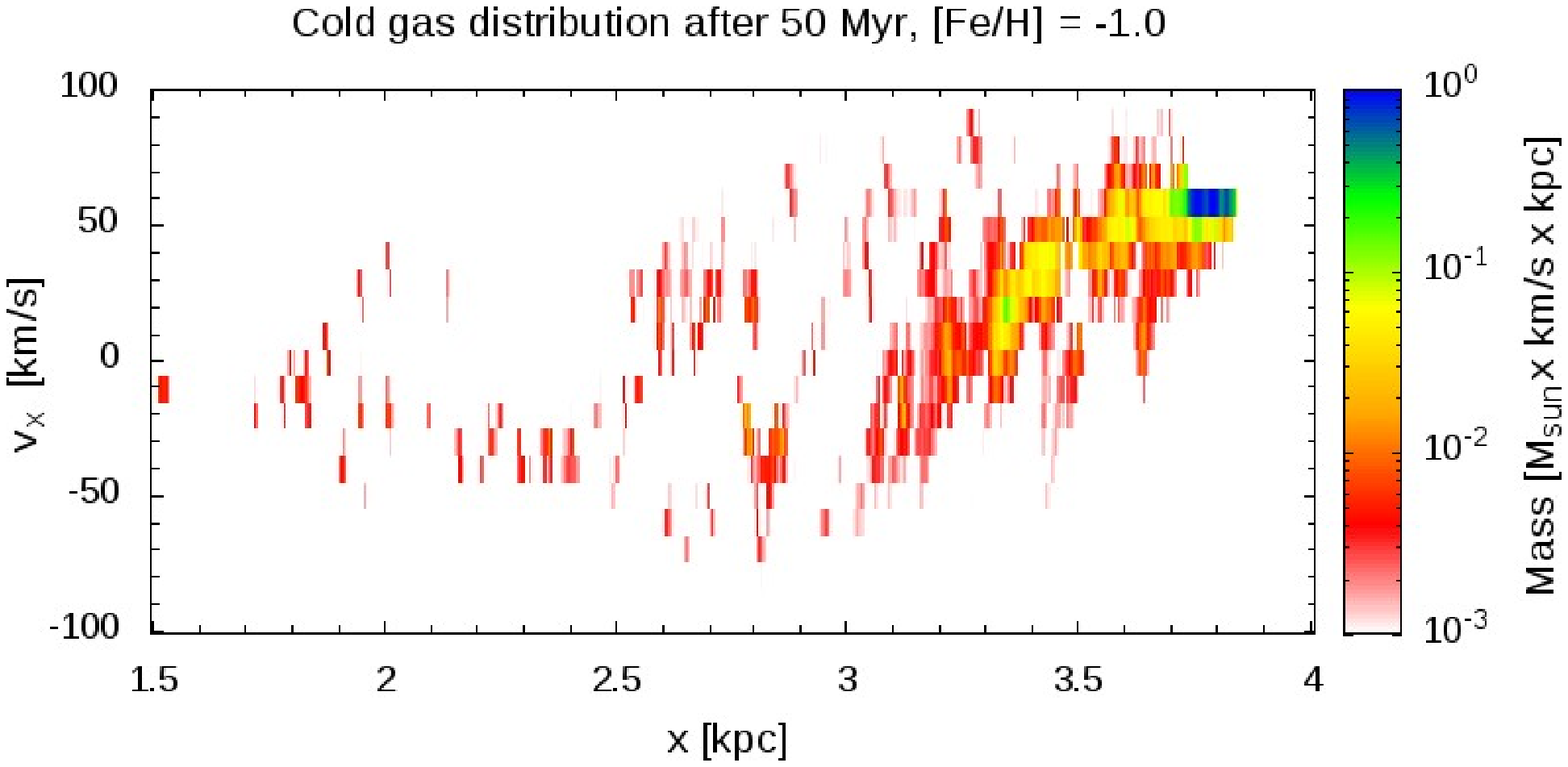,width=.9\hsize}}
\caption{Distribution of velocities as a function of distance down the wake
in T\_3\_Z\_4 at $50\Myr$ when cooling is switched off (upper panel) and on (lower
panel)}.\label{fig:vwidth}
\end{figure}

\fig{fig:vwidth} shows the distribution of cold gas in $x$, the direction
of travel, and in $v_x$ for the high-pressure simulation T\_3\_Z\_4 with and
without cooling. Turning on cooling greatly broadens the extent in both $x$
and $v_x$ of low-density cool gas.  In both simulations, the region of
highest gas density runs from $70\kms$ at the cloud's location down to
$\sim20\kms$ $0.5\kpc$ behind the cloud. Thus hydrodynamics leads to a steep
velocity gradient, $\sim80\kms\kpc^{-1}$ in the part of the wake that will
dominate 21-cm emission. \cite{JinLB} assume that a gradient in the measured
line-of-sight velocity along Complex A, $\sim12\kms\kpc^{-1}$, is entirely
produced by the Galaxy's gravitational field. In light of our simulations
this assumption should be treated with some caution.

The lower panel of \figref{fig:vwidth} shows that the velocity width of the
wake increases with distance from the cloud for $\sim800\pc$ from the cloud
as a result of the lower envelope (velocities in the opposite direction to
the cloud's motion) moving downward towards $-100\kms$, while the upper
boundary remains at $\sim90\kms$. This is further evidence that turbulent
eddies impart peculiar velocities comparable to the cloud's velocity.  Just
behind the cloud material has to flow faster than the cloud in order to flow
into the space vacated as the cloud moves on. Further back similar vortices
carry cold gas away from the cloud with similar velocities.  In the region
$1$ to $2\kpc$ behind the cloud, the turbulence damps quite rapidly.

From these simulations we conclude the following.

\begin{itemize}

\item There is clear evidence that the simulations' limited resolution is
adequate for determining the rate at which gas is stripped from a cloud,
although the two-dimensional nature of the simulations suggests that they
will underestimate the stripping rate by a factor $\sim2$. 

\item The stripping rate
tends to increase with time as a result of the cloud being flattened by ram
pressure.

\item The stripping rate is best determined from simulations in which cooling
is turned off, and it is found to lie close to the rate expected on
dimensional grounds. In \S\ref{sec:theory} we saw that this rate is very
similar to the critical rate at which cooling takes over from evaporation. 

\item Whether stripped gas is evaporated or leads to the condensation of
coronal gas depends on the structure of the turbulent wake. Consequently the
finite spatial resolution of the simulations makes it impossible to determine
with confidence the combinations of pressure and metallicity which divide
evaporation from condensation. However the simulations confirm analytic
arguments, which imply that for a given cloud there is a critical path in the
space spanned by the coronal pressure and metallicity which divides
situations in which, at high pressure or metallicity, coronal gas condenses
in the wake and those in which, at lower pressure or metallicity, the gas stripped from the cloud is evaporated
by the corona. Both the simulations and analytic arguments suggest that the
parameters of the Galactic corona lie close to   this
critical path.

\item The  neutral gas that trails the cloud is strongly influenced
by hydrodynamics forces and cannot be considered to be on an orbit. There
is a large velocity gradient along its high-density ridge in
velocity-position space. Within $\sim2\kpc$ from the cloud the velocity width
of the stream is of order of the cloud's velocity.

\end{itemize}

\section{Discussion}\label{sec:discuss}

The picture developed here of the connection between the Galactic fountain
and accretion onto the star-forming disc differs materially from that
proposed by FB08: in that paper \hi\ clouds grew in mass as they moved
through the corona; here each cloud loses mass, but the mass of cold gas in a
cloud and its wake taken together increases with time. A natural question is
whether the present picture predicts a similar overall picture of 21-cm
emission to that proposed by FB08. If it does, it will fit the data
reasonably well.

FB08 fitted 21-cm data cubes for NGC\,891 and NGC\,2403. Individual clouds
are for the most part not resolved in these data, so the observed emission is
due to many superposed clouds. In the FB08 model, each cloud places a blob of
emission in the cube, centred on its sky-position and line-of-sight velocity,
and smeared by the angular and velocity resolution of the survey. If the
picture developed here is correct, the emission of each cloud is extended in
velocity and is elongated on the sky.  Very sensitive 21-cm data would be
required to see the full extent of the trail of an individual cloud -- the
antenna temperature will drop by two orders of magnitude within $0.5\kpc$ and
$\sim25\kms$ of the cloud's position and velocity. However, the integrated
emission from many individually undetectable trails must contribute to the
\hi\ ``beards'' of nearby galaxies.   In a forthcoming study  we will modify
the pseudo-data cubes of FB08 to include the extended emission from the
trails of clouds. We suspect that these improved data cubes will have similar
observable properties to those of FB08 because the velocity centroid of each
cloud's total emission will be unchanged.

Clearly it is to surveys of our own Galaxy that we must turn for evidence
that clouds have \hi-rich wakes. Maps of the 21-cm emission of individual
high-velocity clouds generally display a tadpole-like structure: the cloud is
elongated and the point of highest surface brightness lies towards one end
\cite[e.g.][]{Bruens01,Westmeier05}.
In the scenario proposed here, tadpole-like structures can be explained 
by the interaction between the cloud and the ambient (coronal) medium. 
This interaction leads to the formation of trailing material behind the cloud.
   
Even in our Galaxy most emission from extraplanar gas is unresolved in the
sense that along any direction in the LAB survey, emission is detected over a
wide band in velocity around 0.  However, a scan through the data cube, one
heliocentric velocity at a time, reveals numerous elongated structures along
which there is a systematic trend in velocity. These could well be the wakes
of relatively massive clouds.

Clouds with masses well below the threshold for detection of their 21-cm
emission can be detected through the absorption lines to which they give rise
in the  ultraviolet spectra of background sources. Absorption-line studies
suggest that large \hi\ complexes are associated with numerous small \hi\
clouds \citep{Richter05}. Could these small clouds be the knots of cold gas
visible in the wake of \figref{fig:coolTplot}? If this interpretation is
correct, the small clouds would be found only on one side of the large
complex, and they would have a mass spectrum that was restricted to masses
very much smaller than that of the complex.

In the simulations the metallicity of the cloud is the same as that of the
corona, whereas  real-world clouds will be more metal-rich than
the corona by a factor up to 10. Whether gas condenses or evaporates depends
on the cooling rate of gas that is roughly a 50--50 mixture of gas stripped
from the  cloud and coronal gas, so the most realistic simulations are those
in which the universal metallicity is about half that of real clouds; that is
the simulations with $\feh\sim-0.5$.

We must obviously ask how valid a guide two-dimensional simulations will be
to the real, three-dimensional problem. Reducing the dimensionality of the
problem reduces the number of high-wavenumber modes relative to the
immediately driven low-wavenumber modes, thus making the power-spectrum of
turbulence less steep. This argument suggests that turbulent mixing will
be more effective in two dimensions than in three, with the consequence that
our simulations have a tendency to over-estimate the pressure or metallicity
of the transition from evaporation to condensation. However, this conclusion
must be considered very tentative at this time.

\subsection{Relation to prior work}

The ablation of clouds that move through a low-density medium has been
discussed by, among others, \cite{Murray93}, \cite{Dinge97}, \cite{Vieser07}
and \cite{HeitschP09}. In
addition to studying the ablation of pressure-bounded clouds as here, some of
these
authors have considered also gravitationally bound clouds and
included thermal conduction in addition to radiative cooling. \cite{Vieser07}
find that thermal
conduction stabilises a moving cloud by reducing the magnitude of
the velocity gradient in the boundary layer where the cloud meets the ambient
medium; a smaller velocity gradient leads to slower growth of the
Kelvin-Helmholtz instability. 

Most previous simulations assume the flow to be axisymmetric around the
velocity of the cloud's motion. In such simulations material becomes trapped
on the assumed symmetry axis, where the radial velocity must vanish by
symmetry.  \cite{Murray93} also simulated clouds with the symmetry assumed
here and found that the main results were independent of the adopted
geometry. 

\cite{Murray93} found ablation to proceed faster than we do and interpreted their
ablation timescale as the inverse of the Kelvin-Helmholtz growth rate, which
is larger than $\alpha_{\rm crit}$ by $\sim(T_\h/T_\c)^{1/2}$. The difference
between their ablation rate and ours probably arises from their use of a
different criterion for identifying cloud gas: they took this to be the mass
within an appropriate sphere, whereas we have defined cloud gas to be cool
gas, regardless of where it resides. In other respects our results are in
accordance with earlier findings.

\rev{The simulations most comparable to ours are those of \cite{HeitschP09}, who
simulated the ablation of cool clouds that fall towards the disc through the
corona. Consequently, their characteristic coronal density,
$n_\h\simeq10^{-4}\cm^{-3}$, was a factor 2 to 4 lower than ours. Their
simulations used a three-dimensional grid, so their
grid spacing was coarser and they could not provide evidence of numerical
convergence. Their clouds, which had similar initial masses to ours,
fragmented in a similar fashion. In most simulations the total mass of \hi\
in the computational volume declined with time, although in some simulations
an upturn in the \hi\ mass is evident at late times as the cloud approaches
the plane. Our numerical results are entirely consistent with theirs,
although our physical motivation is different and, crucially, our parameter
regime extends to the higher coronal densities expected near the plane.}

\rev{Recent attempts to model extraplanar \hi\ include those of \cite{Barnabe06}
and \cite{Kaufmann06}. \cite{Barnabe06} investigated the equilibria of
differentially rotating distributions of gas in flattened gravitational
potentials. They showed that by making the specific entropy and angular
momentum of the gas vary appropriately within the meridional plane, the
kinematics of the gas can be made consistent with the data for \hi\ around
NGC\,891.  They noted that dynamical equilibrium required the gas to be too
hot to be neutral, so the gas within their model could not be the observed
as \hi\ itself, but suggested it might be coronal gas within which \hi\ clouds
were embedded as almost stationary structures \cite[see also][]{Marinacci09}.} 

\rev{\cite{Kaufmann06} studied the
settling of hot gas within the  gravitational potentials
of spherical galaxy-sized dark halos. The gas was initially spinning with a
rotation velocity that was independent of radius (so specific angular
momentum $j\sim r$). A prompt cooling catastrophe caused cold gas to
accumulate in a centrifugally supported disc, and clouds of cold gas
were subsequently seen to be falling through hotter gas towards this disc.
This simulation is interesting but one has to worry that the cool clouds are
numerical artifacts. It is well known that standard smooth-particle hydrodynamics
artificially stabilises contact discontinuities \citep{Agertz,Price}.
Moreover, observed high-velocity clouds all have masses that lie below the
resolution limit of the simulations of Kaufmann et al., so the clouds they
see have no counterparts in reality. Finally, \cite{BinneyNF09} showed that
unless the specific entropy profile of coronal gas is unexpectedly flat, a
combination of buoyancy and thermal conductivity (which was not included in
the Kaufmann et al.\ simulations) suppresses thermal instability.}

\section{Conclusions}\label{sec:conclude}

There is abundant evidence that in galaxies like ours, star-formation powers
a fountain that each gigayear carries $\ga5\times10^9\msun$ of \hi\ to
heights in excess of $1\kpc$ above the plane.  Several lines of argument
strongly suggest that galaxies like ours are surrounded by gas at the virial
temperature -- coronae. The density of the corona, which must vary with
position, is very uncertain, especially in the region above the star-forming
disc. However, there are indications that this density is such that the local
coronal cooling time is of order a gigayear. At least half of the baryons in
the Universe are believed to reside in coronae and their extensions to
intergalactic space.

Models of the chemistry and stellar content of the Galactic disc require the
disc to accrete $\ga1\msun\yr^{-1}$ of low-metallicity gas. The corona is the only
reservoir of baryons that is capable of sustaining an infall rate of this
order for a Hubble time. Therefore there is a strong prima-facie case that
star formation in the disc is sustained by cooling of coronal gas.

The dynamical interaction of \hi\ clouds of the fountain with coronal gas is
inevitable. For any plausible coronal density, the ram pressure arising from
motion through the corona leads to non-negligible loss of momentum by
fountain clouds. If this momentum were retained by the corona rather than
returned to the disc, the corona would rapidly become rotation-dominated. We
have not pursued this possibility because we think it is more likely that
coronal gas that absorbs momentum from fountain clouds is shortly thereafter
accreted by the disc. We suggest that the absorption proceeds as follows: (i)
coronal gas strips gas from the leading edge of the cloud as a result of
Kelvin-Helmholtz instability; (ii) in the turbulent wake of the cloud, the
stripped gas mixes with a comparable mass of coronal gas; (iii) as a result
of this mixing the cooling time becomes shorter than the cloud's flight time
and coronal and stripped gas together form knots of \hi\ that trail behind
the cloud and fall onto the disc within a dynamical time. This scenario is
suggested by a combination of analytic and observational arguments, and
supported by hydrodynamical simulations.

Analytic arguments imply that for a given coronal pressure and metallicity
there is a critical rate of mass loss by a cloud, $\acrit$, such that at
lower mass-loss rates, stripped gas will be evaporated by the corona, and the
total mass of \hi\ will decrease during a cloud's flight. By contrast, when
the mass-loss rate exceeds $\acrit$, stripped gas will lead to condensation
of coronal gas, so the mass of \hi\ increases over time. Dimensional
arguments suggest that the actual mass-loss rate must lie close to $\acrit$.

We have used grid-based hydrodynamical simulations of the flight of a cloud
to check the analytic arguments. Any hydrodynamical simulation is severely
limited by its finite spatial resolution. However, we present evidence from
simulations in which radiative cooling has been switched off that our
simulations have sufficient resolution to provide reliable estimates of the
mass-loss rate. This rate is such that clouds with masses $\lta 1000\msun$
(eq.~\ref{eq:mcrit}) will be totally disrupted before they return to the
disc.

Simulations that include radiative cooling confirm the existence of a
critical mass-loss rate $\acrit$ that depends on coronal pressure and
metallicity in the expected manner. On account of the restricted resolution
of our simulations, we cannot give a definitive value for $\acrit$ within the
local corona. However, we can argue that if the critical mass-loss rate lay
above the actual mass-loss rates, the coronal density and metallicity would
rise secularly at the expense of the star-forming disc. As a consequence, the
$\acrit$ would decrease until it fell below the actual mass-loss rate. That
is, the conditions at the base of the corona have a tendency to adjust until
they lead to accretion by the disc.

Although the present model differs materially from that of FB08 in that we
envisage the masses of clouds decreasing rather than increasing over time, it
seems likely that when the model is used to simulate data cubes for the
galaxies studied by FB08, similar agreement with the data will be achieved.
This is because FB08 correctly give the dependence on time of both the mass of
\hi\ associated with a given cloud, and its velocity centroid. \rev{
However,
a revision of the FB08 models is required to test this conjecture.}

The simulations make detailed predictions for what should be seen in studies
of the high-latitude \hi\ distribution in the Galaxy. Trails behind clouds
should show large gradients in mean velocity that are dominated by hydrodynamical
rather than gravitational forces. High-sensitivity data should reveal small
quantities of gas distributed around the mean velocity by of order the
cloud's velocity. The trail should be studded by knots of cold gas. 

The idea that star-forming discs reach up into the surrounding corona and
grab the relatively pristine gas required to sustain their star formation for
a Hubble time, makes a good deal of sense cosmologically by explaining how
discs can remain star-forming as long as they are not disrupted by a
major merger. If after a major merger there are significant streams of cold
gas, this gas can seed a new star-forming disc, but in the absence of cold
seed-gas, star formation ceases because coronal gas can only condense onto a
disc that is already star-forming. \cite{NipotiB07} argued that thermal
evaporation by coronal gas of filaments of cold gas determines whether a
central cusp reforms when an early-type galaxy experiences a major merger.
Similarly, when a late-type galaxy experiences a major merger, thermal
evaporation of filaments of cold gas can prevent the gas disc reforming.

As a halo proceeds up the clustering hierarchy, the effectiveness of thermal
evaporation increases, and at some point a gas disc is prevented from forming
after a major merger. Following this event, there is a dramatic reduction in
the rate at which coronal gas cools, and the density and X-ray luminosity of
the corona increase rapidly. The higher they get, the lower the
chance that at the next major merger a cool-gas filament will reach the
galactic centre and renew the \rev{stellar cusp}. Thus there is a direct connection
between the process discussed here and the stark contrast in the
detectability with  X-rays of the coronae of star-forming and early-type
galaxies \citep{Rasmussen09}.

The progressive reduction with cosmic time
in the typical halo mass of galaxies making the  transition from the blue
cloud to the red sequence (``downsizing''), follows from the
decrease in the abundance of cold infalling gas with both cosmic time and
halo mass-scale that is expected on analytic grounds and observed in
ab-initio cosmological simulations. It would be interesting to add
this idea to semi-analytic models of the evolution of the galaxy population.

\section*{Acknowledgments}

We thank an anonymous referee for his/her valuable comments.
Most of the numerical simulations were performed using the BCX system at
CINECA, Bologna, with CPU time assigned under the INAF-CINECA agreement
2008-2010. FM gratefully acknowledges support from the Marco Polo
program, University of Bologna.

\label{lastpage}

\end{document}